\documentclass[conference]{IEEEtran}
\usepackage{color,xcolor}
\usepackage{subfigure}
\usepackage{psfrag}
\usepackage{cite}
\usepackage{booktabs}
\usepackage{amsmath}
\usepackage[hidelinks=true,bookmarks=false]{hyperref}

\usepackage{multirow,multicol}
\usepackage{epsfig}
\usepackage{epstopdf}
\usepackage{balance}
\usepackage{mathtools, cuted}
\usepackage{lipsum}
\usepackage{tabularx}
\usepackage{booktabs} 
\usepackage{enumerate}
\usepackage{graphicx,cite}
\usepackage{times}
 \usepackage{amssymb,lipsum}
 \usepackage{algorithm,algpseudocode,graphicx}
\usepackage{mathtools}

\begin{document}
\bstctlcite{IEEEexample:BSTcontrol}
\title{A Novel Joint DRL-Based Utility Optimization for UAV Data Services}
\author{
\IEEEauthorblockN{Xuli Cai, Poonam Lohan, and Burak Kantarci}
\IEEEauthorblockA{\textit{University of Ottawa, Ottawa, ON, Canada}\\
\{xcai049, ppoonam, burak.kantarci\}@uottawa.ca}
\vspace{-0.3in}

}
\maketitle
\begin{abstract}In this paper, we propose a novel joint deep reinforcement learning (DRL)-based solution to optimize the utility of an uncrewed aerial vehicle (UAV)-assisted communication network. To maximize the number of users served within the constraints of the UAV's limited bandwidth and power resources, we employ deep Q-Networks (DQN) and deep deterministic policy gradient (DDPG) algorithms for optimal resource allocation to ground users with heterogeneous data rate demands. The DQN algorithm dynamically allocates multiple bandwidth resource blocks to different users based on current demand and available resource states. Simultaneously, the DDPG algorithm manages power allocation, continuously adjusting power levels to adapt to varying distances and fading conditions, including Rayleigh fading for non-line-of-sight (NLoS) links and Rician fading for line-of-sight (LoS) links. Our joint DRL-based solution demonstrates an increase of up to 41\% in the number of users served compared to scenarios with equal bandwidth and power allocation.
\end{abstract}

\section{Introduction}
Uncrewed Aerial Vehicles (UAVs), commonly known as drones, have emerged as a promising solution for enhancing communication networks, especially in areas with limited infrastructure or during emergency situations \cite{PARVARESH2022100474}. Many existing works 
\cite{9847301, 9916390, 8689172, 10517509} have proposed various UAV applications in emergency situations. The research work \cite{9916390} presents a system for emergency control using UAVs, offering a method to determine the optimal UAV fleet size based on reliability and task requirements. This approach enhances UAV deployment in emergencies by considering redundancy and aircraft quality. The findings enable efficient UAV configuration for emergency operations. The agility, flexibility, and cost-effectiveness of UAVs make them ideal for temporary or supplementary network coverage. The study in \cite{8689172} examines the deployment of UAVs as aerial base stations (ABSs) to restore communication networks following severe disasters. The work in \cite{10517509} presents an architecture to integrate UAVs with Non-Terrestrial Networks (NTN) for quality-aware and sustainable data collection from the ground. However, integrating UAVs into communication networks poses challenges in efficient power and bandwidth management.

Power allocation in UAV-assisted networks is crucial for ensuring prolonged operation and coverage. UAVs have limited energy resources, and optimizing their power usage is essential for maximizing their utility and communication capabilities. The work in \cite{8514812} aims to maximize the rate of a device-to-device (D2D) pair for a downlink UAV-aided wireless communication system, where D2D users coexist in an underlying manner. The research paper \cite{8897110} addresses the UAV utility maximization problem while providing the stochastic analysis of rate coverage probability. However, the consideration of practical air-to-ground channel modeling with LoS link and NLoS link incorporating small-scale fading is not considered.  The research in \cite{8928003} analyzes the use of a UAV as a relay node in emergency communications, deriving a closed-form expression for outage probability in Rician fading channels. It presents a convex optimization problem for minimizing the outage probability with a power constraint, and simulation results confirm the efficacy of the solution over traditional methods. The paper \cite{8379427} addresses an uplink power control issue in UAV-assisted wireless communications, optimizing multiple factors such as UAV altitude, antenna beamwidth, location, and ground terminal resources to minimize total uplink power while meeting rate requirements. These papers primarily discuss traditional tools for optimization which can be difficult to adapt to varying traffic demands and practical considerations in wireless channel modeling.

Recent advances in DRL provide promising solutions for resource allocation in dynamic environments. These algorithms learn optimal policies through environmental interactions, making them ideal for complex scenarios. The work in \cite{9952663} seeks to maximize the UAV's service time and downlink throughput by utilizing the DDPG algorithm. The study in \cite{parvaresh2023continuous} presents an actor-critic network on a continuous action space for optimal  UAV deployment so that it covers the maximum possible number of users with the highest possible sum data rate. Additionally, in work\cite{10033863}, a comparative analysis is conducted to assess the strengths and limitations of employing DQN and DDPG algorithms in autonomous decision-making for UAVs. Leveraging the discrete action capabilities of DQN \cite{mnih2013playing} and the continuous action handling of DDPG \cite{lillicrap2019continuous}, resource allocation to the users can be efficiently optimized which in turn maximizes UAV's utility.  Thus, primary contribution of our work is three-fold: 

1) A novel joint DRL-based algorithm is proposed for optimal resource allocations to the users to maximize UAV utility for heterogeneous data services. DQN manages discrete bandwidth resource block allocation, while DDPG continuously adjusts power levels. 

2) A practical air-ground channel modeling with LoS links having Rician fading and  NLoS links having Rayleigh fading has been considered. 

3) A trained DQN model optimizes resource block allocation for users based on power levels determined by DDPG algorithms. This hybrid approach enhances overall system performance, increasing the number of served users by 41\% compared to equal bandwidth and power allocation scenarios.

The rest of the paper is structured as follows: Section II delineates the system model and problem formulation. Section III elaborates on the proposed joint DRL-based solution. Section IV presents the numerical findings, while Section V encapsulates the conclusions drawn from this study. 


\section{System Model and Problem Formulation}
\subsection{System and Channel Models}
We investigate a UAV-assisted communication system, where $N$ ground users (GUs) are uniformly distributed across a two-dimensional circular field $\Psi$ with a radius of $R$ meters. In this setup, illustrated in Fig. 1, these users communicate with a UAV that remains static and hovers at an altitude of $h$ meters above the center of $\Psi$. Each ground user $GU_i$ is identified by $i\in I{\triangleq}{1,2,\ldots,N}$, and their respective positions are defined by coordinates $(x_i,y_i,0)$ relative to the center of $\Psi$, $(0,0,0)$. The UAV's location is represented by coordinates $(0,0,h)$ in three-dimensional space.

Consider a designated user $GU_i$ depicted in Fig. 1, situated at a distance $r_i\triangleq\sqrt{x_i^2+y_i^2}$ from the center of $\Psi$ and the elevation angle of the UAV to that user is $\theta_i$ rad. For simplicity, we utilize Euclidean distance metrics in our analysis. Given that the UAV maintains an altitude of $h$ meters above the center of $\Psi$, the distance between user $GU_i$ and the UAV can be calculated as $d_i\triangleq \sqrt{r_i^2+h^2}= \frac{h}{sin(\theta_i)}$. Both the UAV and all users are assumed to be equipped with a single antenna.
\begin{figure}[t] \centering
    \includegraphics[width=8.7cm]{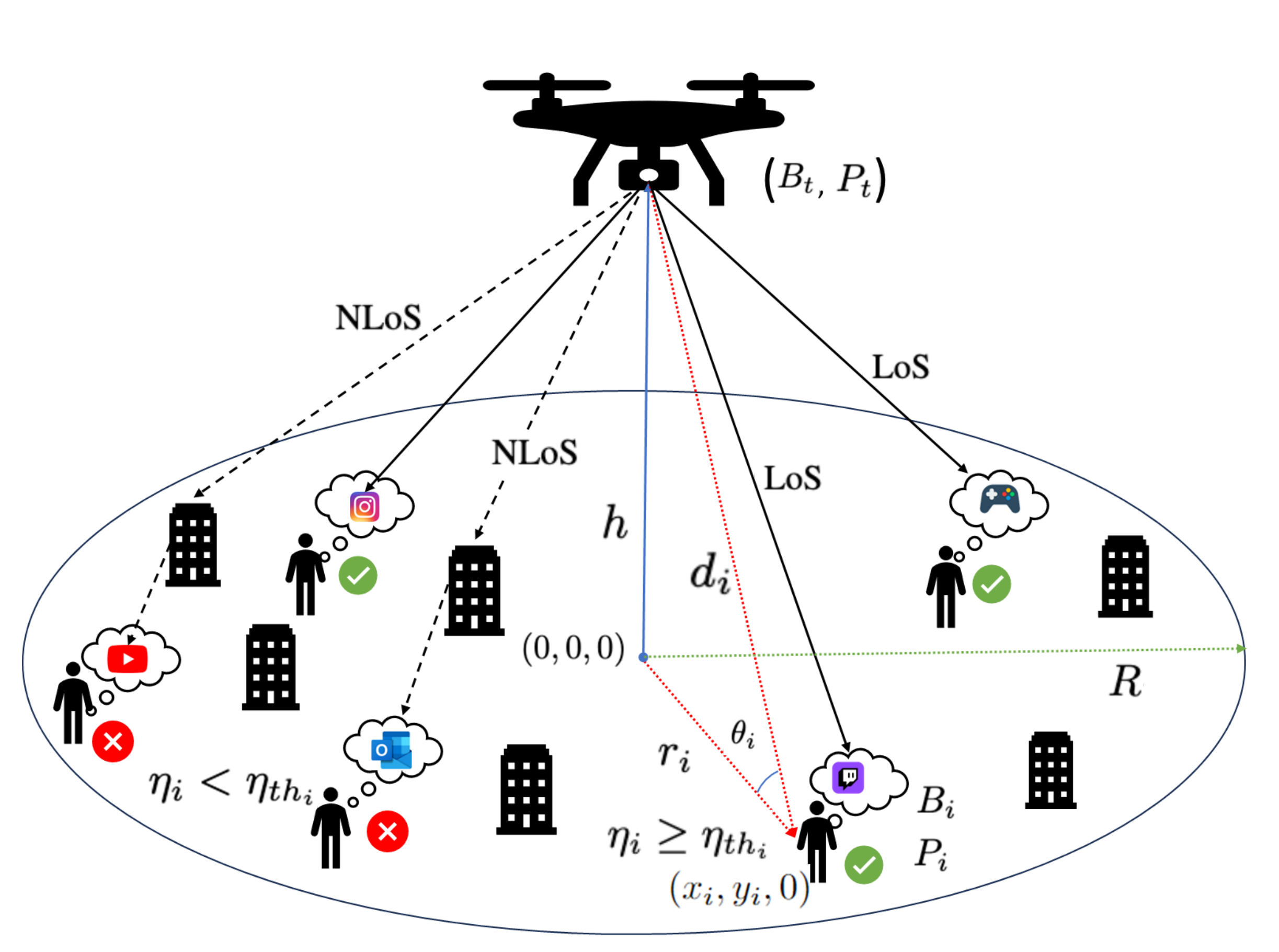} 
    \caption{System model representing ground users in circular field served by a UAV}
    \label{fig:sys_mod}
\end{figure}

 One common approach for air-to-ground channel modeling between the UAV and users is to consider the LoS and NLoS links separately along with their different occurrence probabilities \cite{Hourani}. Note that for NLoS link, the path loss exponent factor $\alpha_{NLoS}$ is higher than that in the LoS link $\alpha_{LoS}$ due to the shadowing effect and reflection from obstacles.  Also to incorporate the effect of small-scale fading, we are considering Rician fading in LoS links and Rayleigh fading in NLoS links. Consequently, the random channel power gains, $g_i$, for LoS link are noncentral-$\chi^2$ distributed with mean $\mu$ and rice factor $K$ \cite{simon}, and the random channel power gains, $k_i$, for NLOS link are exponentially distributed with mean $\mu$. Here,  $\mu$ is the average channel power gain parameter that depends on antenna characteristics and average channel attenuation.  With this consideration, the received signal-to-noise ratio (SNR) for Los and NLos links at  $GU_i$ can be written as:

\begin{equation}
    \text{SNR}_{i,\text{LoS}} = \frac{P_i g_i d_i^{-\alpha_{LoS}}}{B_i\sigma^2}, \; \forall i\in I,  \quad \text{LoS link}, 
\end{equation}
\begin{equation}
    \text{SNR}_{i,\text{NLoS}} = \frac{ P_i k_i d_i^{-\alpha_{NLoS}}}{B_i\sigma^2}, \; \forall i\in I,  \quad \text{NLoS link},
\end{equation}

where $P_i$ and $B_i$, respectively, are the transmission power and bandwidth allocated to user $GU_i$. $\sigma^2$ denotes AWGN (additive white Gaussian noise) power density. The probability of LoS link between  $GU_i$ and UAV depends upon the elevation angle $\theta_i=\sin^{-1}(\frac{h}{d_i})$, density and height of buildings, and environment. The LoS probability $p_{LoS}$ is written as \cite{Hourani}:
\begin{equation}\label{eq:plos}
p_{LoS}={1}/{(1+C\exp(-B[({180}/{\pi})\theta_i-C]))},
\end{equation} 
where $C$ and $B$ are constants that depend on the environment (rural, urban, dense urban). The probability of NLoS link is $p_{NLoS}=1-p_{LoS}$. Thus the effective SNR received by user $GU_i$ is expressed as:
\begin{equation}\label{eq:eq1}
    \text{SNR}_{\text{eff}_i} = P_{\text{LoS}_i} \cdot \text{SNR}_{\text{LoS}_i} + P_{\text{NLoS}_i} \cdot \text{SNR}_{\text{NLoS}_i}
\end{equation}
Using Shannon's capacity formula and SNR representation from equation \eqref{eq:eq1}, the data rate, $\eta_i$ bits per sec  (bps) for user $GU_i$ through UAV communication link  can be expressed as:
 \begin{align}\label{eq:eq2}
\eta_i\triangleq B_i \log_2(1+\text{SNR}_{\text{eff}_i})  \; \forall i\in I.
\end{align}

Note that a  user $GU_i, \forall i\in I$, is considered under UAV rate-coverage or served by the UAV, if the data rate for that user is greater than its desired rate threshold $\eta_{th_i}$, i.e., $\eta_i\geq \eta_{th_i}$.

\subsection{Problem Formulation}
Following the system model, our objective is to enhance the UAV utility by maximizing the number of served users, $N_s$. This objective can be achieved by optimally allocating the bandwidth and power resources to users, while considering the heterogeneous data rate demand, $\eta_{th_i}$, for each user $GU_i,\; \forall i\in I\triangleq\{1,2,\ldots, N\}$, limited power budget $P_t$, and bandwidth resources $B_t$ constraints of the UAV. So, the proposed design framework is mathematically expressed as follows:
\begin{align}
&(\mathcal{P}):\;\underset{N_s,\bold{P},\bold{B}}{\text{maximize}}\; N_s,\;\\
 &\text{s.t.:}\; (C1):  \eta_i(P_i,B_i) \geq \eta_{th_i},\; \forall i\in (1,2,...N_s),\nonumber \\ 
 &(C2): \underset{i=1}{\overset{N_s}{\sum}}P_i\leq P_t, \nonumber \;
 (C3): \underset{i=1}{\overset{N_s}{\sum}} B_i\leq B_t,\;
 (C4): N_s\in \text{I},\nonumber\\
 &(C5): P_i\geq 0,\; \forall i\in (1,2,...N_s),\nonumber\\ 
 &(C6): B_i\geq 0,\; \forall i\in (1,2,...N_s).\nonumber
\end{align}
where, constraint $(C1)$ represents the different data rate requirements for each user, $(C2)$  ensures that the sum of power allocated to the served users should not be more than the total power budget, $(C3)$ is the bandwidth resource constraint, $(C4)$ Specifies that the maximum number of served users must be an integer, while $(C5)$ and $(C6)$ are the boundary conditions for power and bandwidth allocation to each user. The power and bandwidth allocation vectors are represented by $\bold{P}= \{ P_1,P_2,...,P_{N_s}\}$ and $\bold{B}=\{ B_1,B_2,...,B_{N_s}\}$, respectively.
The objective, $N_s$, of this optimization problem $\mathcal{P}$ is an integer variable. All constraints and sizes of vectors $\bold{P}$ and $\bold{B}$ are dependent on $N_s$, which itself is unknown. Therefore, $\mathcal{P}$ is a combinatorial, non-convex and NP-hard problem. To solve this problem, we present a novel joint DRL-based algorithm in the next section.

\begin{algorithm} [!h]
\fontsize{9}{9}\selectfont
\caption{Proposed Joint DRL-based Algorithm}
\begin{algorithmic}[1]
\State Initialize ground users positions $(x_i,y_i,0)$ and data rate requirements $\eta{th_i}$ for all $i={1,2...N}$.
\State Initialize DQN model with action-value function $Q$ with random weights $\theta$
\State Initialize replay buffer $D$
\For{episode = 1:J}
    \State Observe initial state $s_1$ = ($P_i$,$B_i$, $(x_i,y_i)$)
    \For{t = 1:T}
        \State With probability $\epsilon$ select a random action $a_t$ from adding one bandwidth block or decreasing one bandwidth block 
        \State Otherwise select $a_t = \arg\max_a Q(s_t, a; \theta)$
        \State Execute action $a_t$ in the UAV network and observe reward $r_t$ = $\eta_i$/$\eta{th_i}$ and update new state $s_{t+1}$
        \State if $r_t>1$ calculate the squared deviation of the ratio $(r_t-1)^2 $ as the penalty to avoid too much resource wastage
        \State Store transition $(s_t, a_t, r_t, s_{t+1})$ in $D$
        \State Sample a random minibatch of transitions $(s_j, a_j, r_j, s_{j+1})$ from $D$
        \State Set $y_j = r_j + \gamma \max_{a'} Q(s_{j+1}, a'; \theta)$
        \State Perform a gradient descent step on $(y_j - Q(s_j, a_j; \theta))^2$ with respect to $\theta$
    \EndFor
    \State Gradually reduce $\epsilon$ (exploration rate)
\EndFor

\State Initialize DDPG model with critic network $Q(s,a|\theta^Q)$ and actor $\mu(s|\theta^\mu)$ with weights $\theta^Q$ and $\theta^\mu$
\State Initialize target network $Q'$ and $\mu'$ with weights $\theta^{Q'} \leftarrow \theta^Q$, $\theta^{\mu'} \leftarrow \theta^\mu$
\State Initialize replay buffer $D'$
\For{episode = 1:K}
    \State Initialize a random process $\mathcal{N}$ for action exploration
    \State Receive initial observation state: $s_1$ = $P_1$,$P_2$...$P_N$,$B_1$,$B_2$...$B_N$ 
    \For{t = 1:T}
        \State Select action $a_t = \mu(s_t|\theta^\mu) + \mathcal{N}_t$ according to the current policy and exploration noise
        \State Execute action $a_t$ (continuous power change for all GUs)
        and use the well-trained DQN model by entering the ($P_i$,$B_i$) state, then get the optimal bandwidth $B_i$,
        at last observe reward $r_t$ = $N_s$ and new state $s_{t+1}$
        \State Calculate the exceed power ($\sum_{i=1}^{n} P_i - P_t $)*10 as the penalty to avoid overusing
        \State Store transition $(s_t, a_t, r_t, s_{t+1})$ in $D'$
        \State Sample a random minibatch of $W$ transitions $(s_i, a_i, r_i, s_{i+1})$ from $D'$
        \State Set $y_i = r_i + \gamma Q'(s_{i+1}, \mu'(s_{i+1}|\theta^{\mu'})|\theta^{Q'})$
        \State Update critic by minimizing the loss: $L = \frac{1}{W}\sum_i(y_i - Q(s_i, a_i|\theta^Q))^2$
        \State Update the actor policy using the sampled policy gradient:$ \nabla_{\theta^\mu}J
        \approx  ({1}/{W})\sum_i \nabla_a Q(s, a|\theta^Q)|_{s=s_i, a=\mu(s_i)} \nabla_{\theta^\mu}\mu(s|\theta^\mu)|_{s_i} $
        \State Update the target networks:
        $ \theta^{Q'} \leftarrow \tau\theta^Q + (1 - \tau)\theta^{Q'} $; \;
        $ \theta^{\mu'} \leftarrow \tau\theta^\mu + (1 - \tau)\theta^{\mu'}$
    \EndFor
\EndFor
\end{algorithmic}
\end{algorithm}

\section{Proposed Joint Model}
Here, we introduce a joint DRL-based algorithm devised for optimizing resource allocation to users and maximizing UAV utility. In this algorithm, the DQN model is employed to allocate optimal bandwidth to each user, considering its allocated power value, channel condition, and path loss effects. Subsequently, based on the allocated bandwidth and channel conditions, DDPG allocates power to users. This iterative process continues until achieving optimal resource allocation, maximizing the number of users served within the given bandwidth and power resources of the UAV. The proposed joint DRL-based algorithm is provided in Algorithm 1. Now the implementation details of DQN and DDPG models are discussed as follows:
 \subsection{DQN Part Implementation}

DQN is a robust DRL method that integrates the traditional Q-learning algorithm with deep neural networks. DQN is particularly effective in environments with discrete action spaces, making it ideal for applications such as discrete bandwidth resource block allocation to the users. It is worth noting that the available bandwidth is partitioned into resource blocks, each consisting of 1.6KHz bandwidth. This approach enables efficient exploration and exploitation of the action space and provides the means to handle complex decision-making processes in dynamically changing environments.

\subsubsection{Model Objective}
The objective of the DQN model is to allocate optimal bandwidth resource blocks to users for various scenarios in the shortest possible time. 

\subsubsection{Initialization}
At the beginning of the training process, uniformly distributed ground users' locations are generated within a defined circular field. Then, the data requirement for each user is randomly generated to simulate different data rate requirements of different applications.

\subsubsection{State and Observation Space}
In this DQN implementation, we initialize the state and observation space with three key elements which are random power levels $P_i$ within a specified limit, random $(GU_i)$ location information ($x_i, y_i,0$), and initial bandwidth $B_i$ for each user.
 Note that, every training episode considers a user with a different initial state.

\subsubsection{Action Space}
The action space in our model consists of two discrete actions: first, addition of one bandwidth resource block and second, subtraction of one bandwidth resource block.
These actions allow DQN model to incrementally adjust bandwidth allocation to each ground user in discrete manner.

\subsubsection{State Updates}
In the step updation, we perform the following steps:
First, the selected bandwidth change is applied to the current bandwidth and the resulting data rate $\eta_i$ is calculated for the user based on the updated bandwidth. Then, the user's data rate is compared to the required data rate $\eta_{th_i}$.

\subsubsection{Reward Function}
The reward is calculated as the ratio of the achieved data rate to the required data rate. Specifically:
$\text{Reward} = {\eta_i}/{\eta_{th_i}}$.

\subsubsection{Penalty Function}
If the reward is over 1, a penalty, $({\eta_i}/{\eta_{th_i}}-1)^2$  would be charged for the reward's deviation from 1. This means that rewards closer to 1 incur smaller penalties, while rewards further away from 1 incur larger penalties. The goal is to minimize resource waste by encouraging rewards that are close to the target value of 1.

\subsubsection{Training Termination}
One training episode is considered completed and is stopped if the reward for a user is between 1 and $1+\epsilon$. Here, $\epsilon$ depends on the data rate gap above the required threshold with an additional resource block while without this additional resource block, $\eta_i$ is less than $\eta_{th_i}$. DQN model training is terminated when the number of steps required to get optimal bandwidth allocation in each episode is converged. This indicates that the model has learned to allocate the optimal bandwidth to meet the user's data rate requirements with the minimum number of time steps.

\subsubsection{Integration with DDPG Training}
While the DQN model provides an efficient method for determining the best bandwidth allocation, continuous power levels pose a challenge for subsequent DDPG training. To address this, we adopt a comparison strategy where power levels are considered identical if they match up to four decimal places. This approximation ensures that the discrete experiences from the DQN model are compatible with the continuous action space of power levels resulting from DDPG model.

\subsection{DDPG Part Implemention}
DDPG is a DRL algorithm that combines Deep Learning and Policy Gradient methods. It works well with continuous action spaces, making it suitable for tasks like UAV network optimization where actions, such as power allocation to users, are not discrete. DDPG uses a deterministic policy to choose actions and a Q-function (critic) to evaluate the expected return of those actions. This combination allows for efficient exploration and stable learning of optimal policies.
\subsubsection{Model Objective}
The objective of the DDPG model is to optimize the power levels for all ground users in a UAV network to maximize the number of users served within the available power budget. This model operates in a continuous action space, making it suitable for fine-tuning power levels.

\subsubsection{State and Observation Space}
The state and observation space are defined as follows: power levels $\bold{P}=(P_1, P_2,....P_N$ and bandwidth allocations $\bold{B}=(B_1, B_2,....B_N$) for all ground users.
These elements capture the current configuration of the network, providing the necessary context for decision-making.

\subsubsection{Action Space}
The action space in the DDPG model consists of continuous adjustments to the power levels for all ground users. This allows for precise control over the power allocation, enabling more efficient resource optimization.

\begin{table}
\centering
\caption{Environmental Parameters Used in the Simulation}
\label{tab:parameters}
\begin{tabular}{@{}cll@{}}
\toprule
\textbf{Symbol}         & \textbf{Description}                      & \textbf{Value} \\ \midrule
\(R\)                   & Radius of the Circular Field            & 200m                      \\
\(h\)                 & Height of the UAV                         & 400m              \\
\(N\)                   & Total Number of Ground Users              & 50                        \\
\(B_{t}\)    & UAV Total Bandwidth                       & 1.6Mhz=1000 blocks            \\
\(P_{t}\)    & UAV Total Power for Transmission          & 1                  \\
\(B\)                   & Environmental Constant (Dense Urban)         & 0.136                     \\ 
\(C\)                   & Environmental Constant (Dense Urban)                   & 11.95                     \\ 
\(\alpha_{\text{LoS}}\) & Path Loss Exponent for LoS                & 2.5                       \\
\(\alpha_{\text{NLoS}}\)& Path Loss exponent for NLoS               & 3.5                       \\
\(\sigma^2\)                 & Noise Power Spectral Density              & 10e-17                    \\ 
$\eta_{th_i}$            & Data Rate Requirement for user $GU_i$  & 1Mbps              \\ 
\(K\)                   & Fading Factor                             & 10                        \\ 
\(\mu\)              & Mean Power                             & 0.5                       \\ 

\bottomrule
\end{tabular}
\end{table}

\subsubsection{State Updates}
In the state updation, the following steps are performed: First, generate a continuous change in the power levels for all users based on the current policy. Second, apply this change to the current state to update the power levels. Third, use the trained DQN model to determine the optimal bandwidth allocation for each user one by one, given the updated power levels. Fourth, If the total used bandwidth exceeds the total bandwidth $B_{t}$, count the number of served users $N_{s}$ as the reward. Fifth, Update the state to reflect the bandwidth allocation by the DQN model for served users. Sixth, Set the bandwidth for unserved users equal to $0$.

\begin{figure*}[t]
\minipage{0.333\textwidth}
  \includegraphics[width=\linewidth]{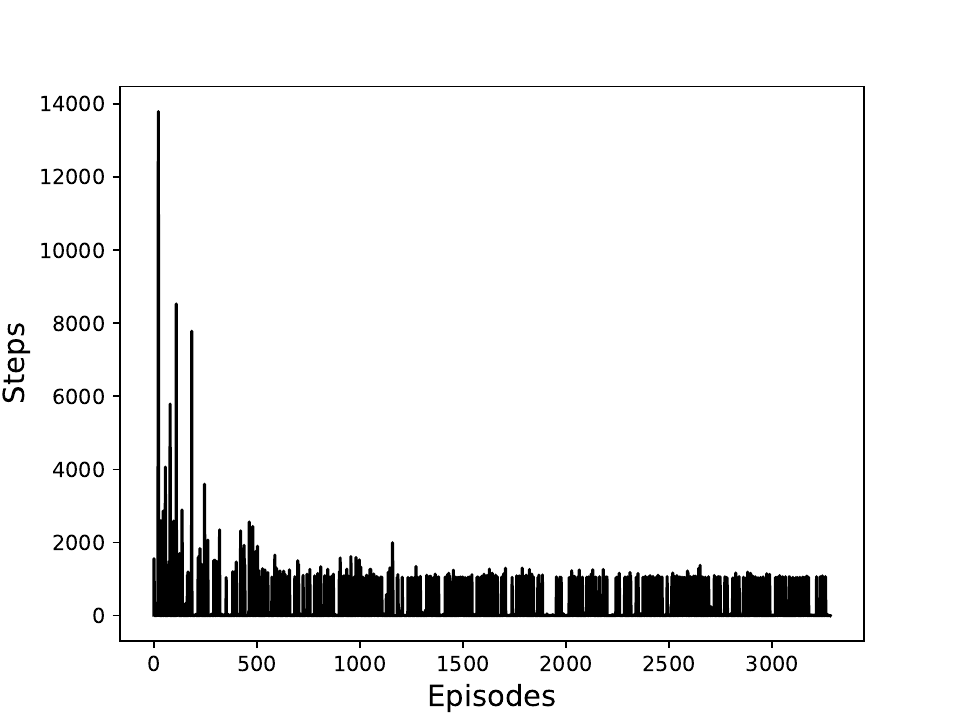}
  \caption{DQN training for bandwidth allocation\\ with random power allocation and user positions}\label{fig:dqn_time}
\endminipage\hfill
\minipage{0.333\textwidth}
  \includegraphics[width=\linewidth]{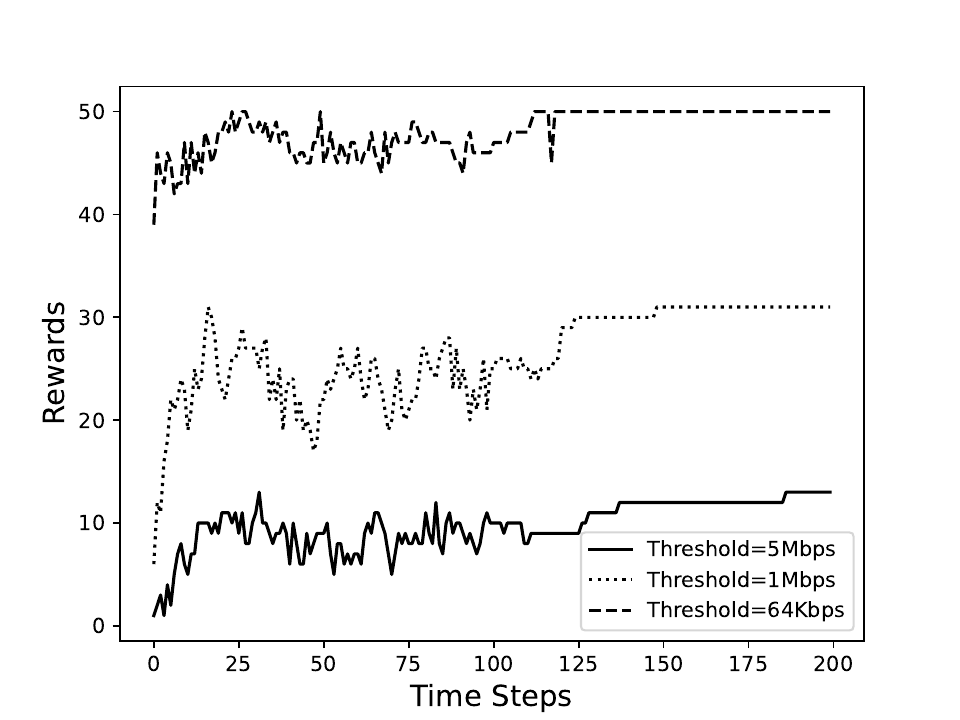}
   \caption{Joint DRL-based model's training\\ with different user thresholds}\label{fig:ddpg_thresholds}
\endminipage\hfill
\minipage{0.333\textwidth}%
  \includegraphics[width=\linewidth]{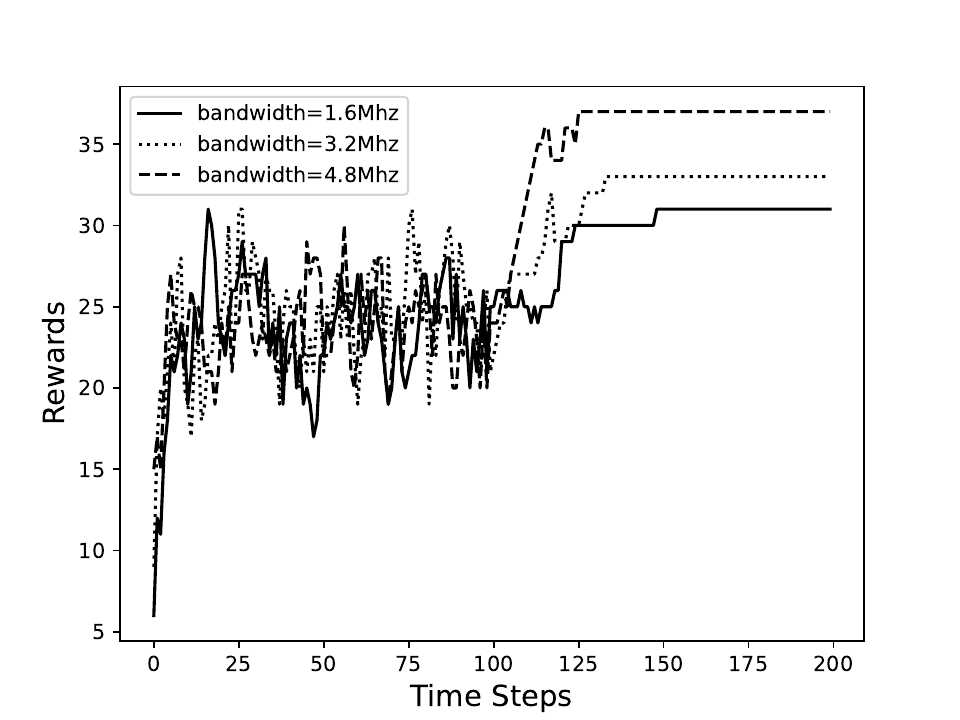}
  \caption{Joint DRL-based model's training with different total bandwidth}\label{fig:ddpg_bandwidth}
\endminipage
\end{figure*}

\begin{figure*}[t]
\minipage{0.33\textwidth}
  \includegraphics[width=\linewidth]
  {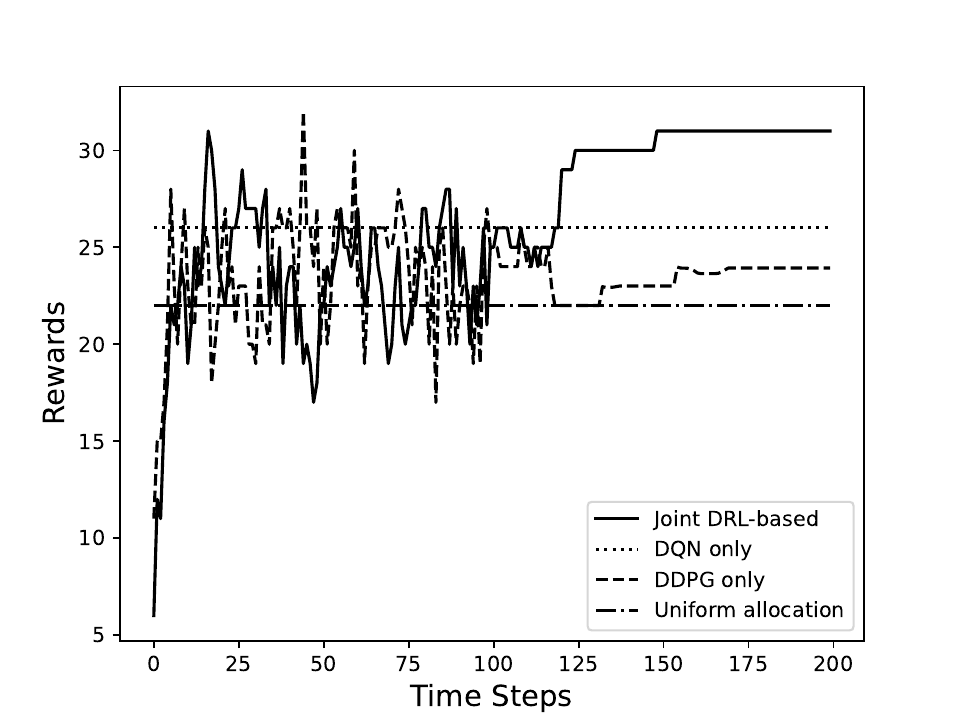} 
    \caption{Comparative analysis of joint\\ DRL-based algorithm performance}\label{fig:ddpg_dqn_comparison}
\endminipage\hfill
\minipage{0.33\textwidth}
  \includegraphics[width=\linewidth] {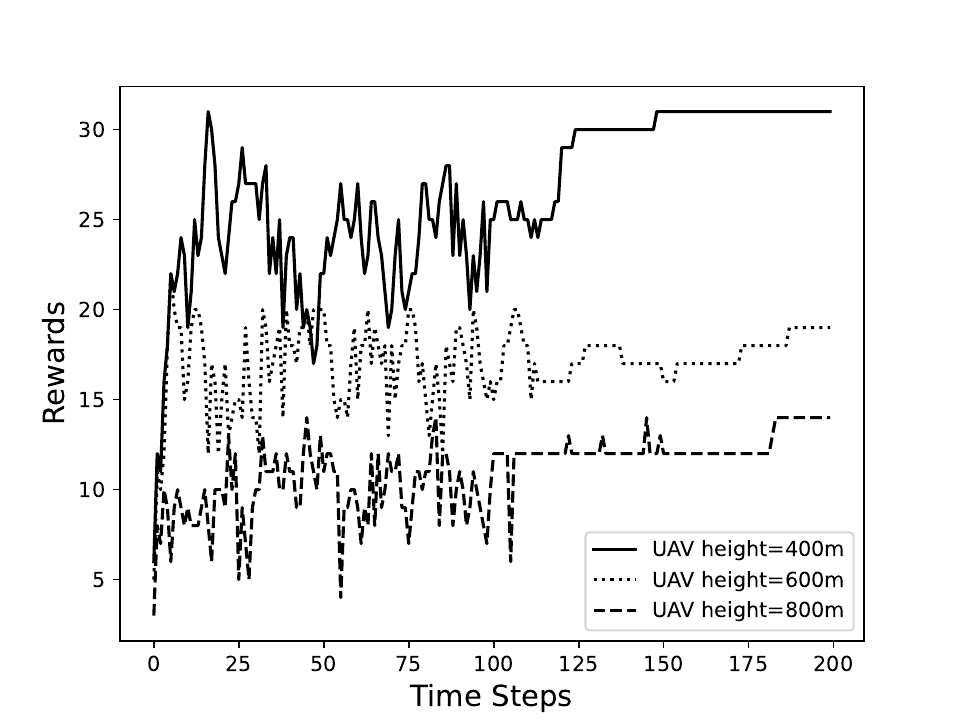} 
    \caption{Joint DRL-based model's training with different UAV heights}\label{fig:ddpg_heights}
\endminipage\hfill
\minipage{0.33\textwidth}%
  \includegraphics[width=\linewidth]{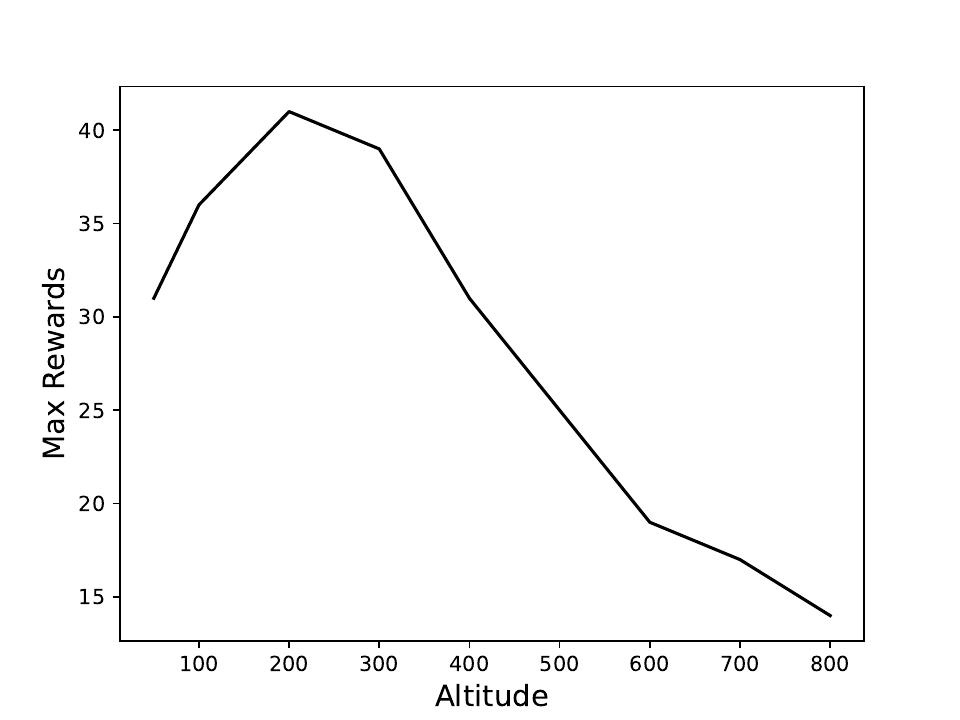} 
    \caption{Rewards at different altitudes}\label{fig:altitude_peak}
\endminipage
\end{figure*}

\subsubsection{Reward Function}
The reward is defined as the number of served users within the total bandwidth and power constraints, i.e.  $\text{Reward} = N_s$. This metric directly incentivizes the model to maximize the efficiency of  power allocation. 
\subsubsection{Penalty Function}
If the sum of power allocated to users exceeds the total power budget, $P_{t}$, a high penalty is charged to the model that is equal to $(\sum_{i=1}^{n} P_i - P_t)*10$. It ensures that the model will learn to avoid using more power than budget.
\subsubsection{Training Termination}
Once the reward stabilizes and converges, we consider the resulting number of users served as the maximum served user number.

\section{Simulation Results and Discussion}
\subsection{Simulation parameters}
Unless otherwise stated all the environmental parameters used in the simulations are provided in Table \ref{tab:parameters}. Regarding hyperparameter tuning of DRL models, the learning rate for DQN is set at 0.0001 to ensure stable convergence by fine-tuning the Q-values accurately, whereas DDPG uses a higher learning rate of 0.001 to learn faster in the continuous action space. Both algorithms share a buffer size of 1,000,000 to store experiences for training, aiding in data decorrelation for stability. The batch size for DQN is 32, which suffices for its simpler Q-value updates, while DDPG uses a larger batch size of 256 for more stable gradient updates in its actor-critic framework. Both use a discount factor $\gamma$ of 0.99 to balance between immediate and future rewards. DQN updates its model every 4 steps to manage computational load and allow Q-values to propagate, whereas DDPG updates at every step for continuous policy and value improvement. Learning starts after 100 steps for both, ensuring enough data in the replay buffer for meaningful training batches. The $\tau$ value for DQN is set to 1, indicating hard updates, while for DDPG, $\tau$ is 0.005, allowing soft updates for gradual and stable target network changes. Both algorithms employ the MultiInputPolicy type for structured input processing, tailored to the problem domain's specific nature. The values of all hyperparameters are summarized in Table \ref{tab:hyperparameters}.

\vspace{-0.5cm}
\begin{table}[H]
\centering
\caption{Hyperparameters for DQN and DDPG}
\label{tab:hyperparameters}
\begin{tabular}{@{}lcc@{}}
\toprule
\textbf{Hyperparameter}      & \textbf{DQN}    & \textbf{DDPG}   \\ \midrule
Learning Rate               & 0.0001          & 0.001           \\
Buffer Size                 & 1000000         & 1000000         \\
Batch Size                  & 32              & 256             \\
Gamma (\(\gamma\))          & 0.99            & 0.99            \\
Train Frequency             & 4               & 1               \\
Learning Starts             & 100             & 100             \\ 
$\tau$ for DQN and DDPG              & 1               & 0.005           \\
Policy                      & MultiInputPolicy       & MultiInputPolicy       \\
\bottomrule
\end{tabular}
\end{table}

\subsection{Results and discussion}
Fig.~\ref{fig:dqn_time} illustrates the convergence time of the DQN algorithm as it searches for  optimal bandwidth allocation strategy. The results indicate that DQN quickly adapts to different user positions and power settings after  500 training episodes, demonstrating its robustness and efficiency in dynamic environments. 

Furthermore, DDPG's performance in the joint DRL-based model utilizing the trained DQN model is extensively tested under various network conditions to assess its adaptability and learning efficiency. In Fig.~\ref{fig:ddpg_thresholds}, the joint DRL-based algorithm's response to different user data rate thresholds is depicted. The training curves suggest that the proposed solution can effectively adjust its policy to meet varying data demands, optimizing bandwidth and power allocation to users to enhance the overall number of users served and fulfill the users' requirements. As discussed earlier, the model begins learning after 100 steps to ensure sufficient data in the replay buffer for meaningful mini-batch sampling and leads to convergence.

Fig.~\ref{fig:ddpg_bandwidth} shows how the proposed joint DRL-based algorithm manages different amounts of total bandwidth. The results indicate a direct correlation between available bandwidth and the network performance and higher bandwidth allows for a higher number of users served. Fig.~\ref{fig:ddpg_dqn_comparison} compares the performance of the proposed joint DRL-based algorithm utilizing DQN for bandwidth allocation and DDPG for power allocation under scenarios where either one of  DDPG and DQN is activated or both are deactivated. In the case when both DRL models are deactivated, uniform/ equal resource allocation is considered in that scenario. It is demonstrated that the proposed solution boosts the served users' number by up to 41\% compared to the equal resource allocation strategy. In the case When only DDPG is active, it represents optimal power allocation alongside equal bandwidth allocation, and when only DQN is active, it represents optimal bandwidth allocation alongside equal power allocation. Compared to these scenarios, our proposed joint DRL-based solution shows $29\%$, and $19\%$ improvements, respectively, in  number of served users.

Fig.~\ref{fig:ddpg_heights} explores the impact of UAV altitude adjustments on joint-DRL model training performance. Higher altitudes generally improve LoS link probability but may introduce higher path loss, which is why with an increase in height, the number of served users decreases. Additionally, Fig.\ref{fig:altitude_peak} illustrates that lowering the UAV height reduces path loss but diminishes LoS link probability. Consequently, there exists an optimal altitude of approximately 200 meters that balances these effects and yields the highest performance.

\begin{figure} 
    \centering
    \includegraphics[width=7cm, height=4cm]{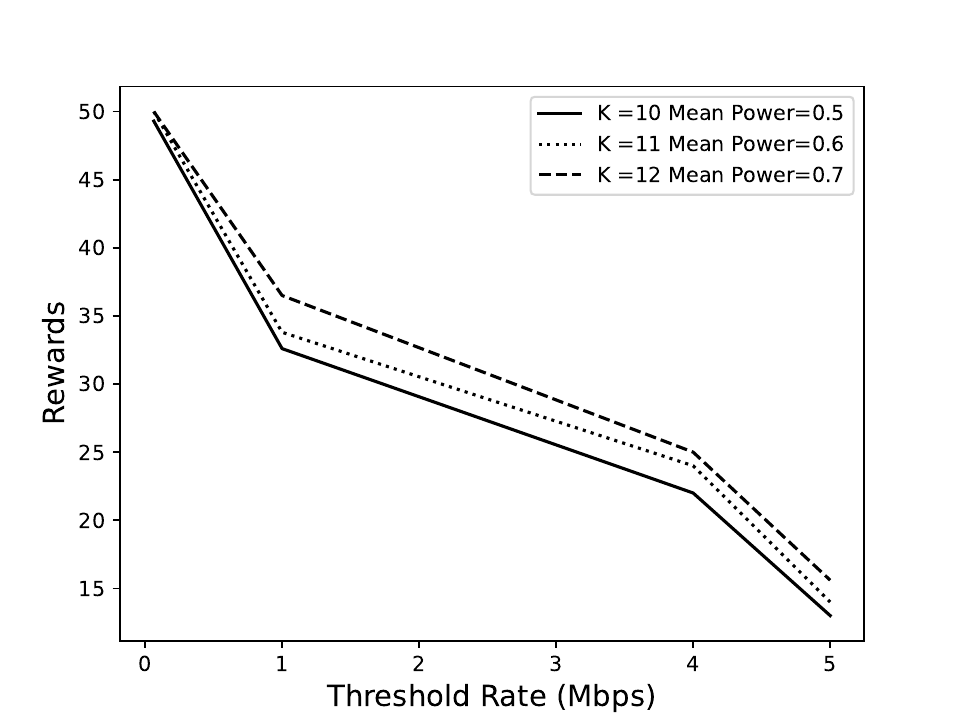} 
    \caption{Rewards under  different total power and thresholds}\label{fig:fading_factors}
\end{figure}

\begin{figure} 
    \centering
    \includegraphics[width=7cm, height=4cm]{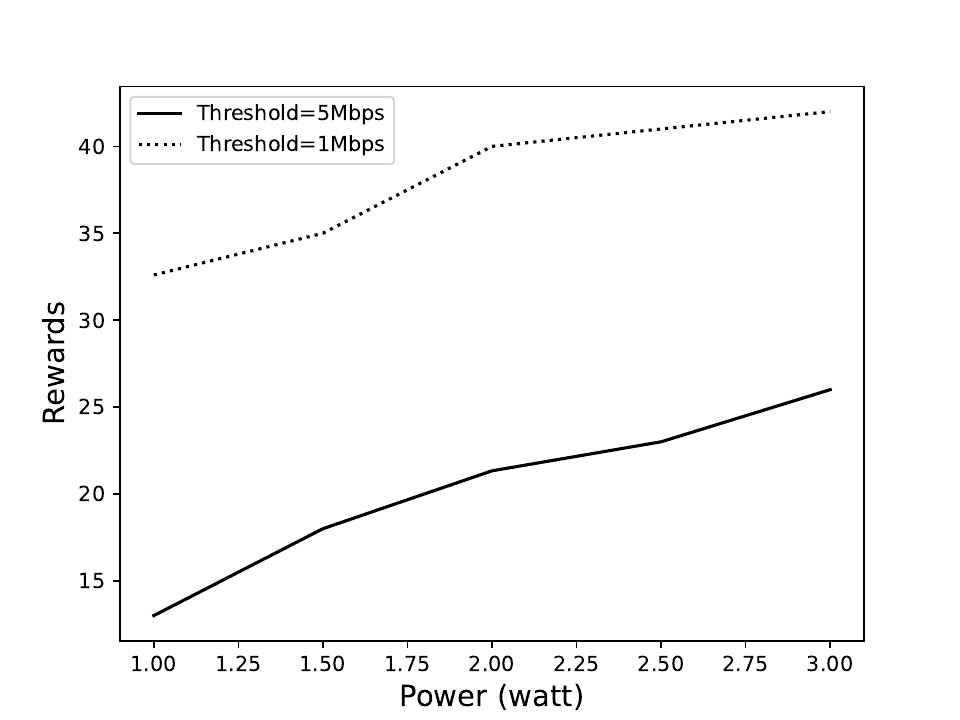} 
    \caption{Rewards under  different total power and thresholds}\label{fig:power_reward}
\end{figure}

Fig.~\ref{fig:fading_factors} shows the number of served users with different $K$ and $\mu$. With better LoS link conditions number of served users is high. Fig.~\ref{fig:power_reward} displays the performance of the proposed joint DRL-based algorithm with the variation in total power budget and users' data rate requirements.  As demonstrated with a higher power budget and lower data rate demands, a larger number of users can be served.

\section{Conclusion}
This study has proposed the amalgamation of two advanced reinforcement learning algorithms, DQN and DDPG, in the resource management of UAV communication networks. A practical air-to-ground channel modeling has been considered while integrating the small-scale fading effects into it. Through a series of simulations, it has been demonstrated how the proposed joint DRL-based algorithm can be effectively utilized to optimize bandwidth allocation and power management in dynamically changing environments and enhance the number of served users by $41\%$ in comparison to the benchmark scheme of uniform resource allocation.  Our ongoing research agenda includes exploration of scenarios involving multiple UAVs, taking into account interfering signals.

.




\vspace{-0.3cm}

\section*{Acknowledgement}
This work is supported in part by the Natural Sciences and Engineering Research Council of Canada (NSERC) CREATE TRAVERSAL program and in part by the NSERC DISCOVERY program.
\vspace{-0.3cm}
\bibliographystyle{IEEEtran}

\end{document}